\documentclass[final,pre,twocolumn,superscriptaddress,floatfix]{revtex4-1}
\usepackage[english]{babel}
\usepackage{latexsym}
\usepackage{amsmath}
\usepackage{color,soul}
\usepackage{hyperref}
\usepackage{fancyhdr}
\usepackage{graphicx}
\usepackage{wrapfig}
\usepackage{hhline}
\usepackage{dcolumn}
\usepackage[english]{babel}
\usepackage{amsmath}
\usepackage{amsfonts}
\usepackage{graphicx}
\begin{document}

\title{ How neutral and niche forces contribute to speciation processes?}
\author{Niccolo Anceschi}
\affiliation{Dipartimento di Fisica e Astronomia G. Galilei and CNISM, INFN, Universit\'a di Padova, Via Marzolo 8, 35131, Padova, Italy}

\author{Jorge Hidalgo}
\affiliation{Dipartimento di Fisica e Astronomia, G. Galilei and CNISM, INFN, Universit\'a di Padova, Via Marzolo 8, 35131, Padova, Italy}


\author{Tommaso Bellini}
\affiliation{Dipartimento di Biotecnologie Mediche e Medicina Traslazionale, Universit\'a degli Studi di Milano, via Fratelli Cervi 93, I-20090 Segrate (MI), Italy}

\author{Amos Maritan}
\affiliation{Dipartimento di Fisica e Astronomia G. Galilei and CNISM, INFN, Universit\'a di Padova, Via Marzolo 8, 35131, Padova, Italy}

\author{Samir Suweis}
\email{suweis@pd.infn.it}
\affiliation{Dipartimento di Fisica e Astronomia G. Galilei and CNISM, INFN, Universit\'a di Padova, Via Marzolo 8, 35131, Padova, Italy}

\begin{abstract}
The evolutionary and ecological processes behind the origin of species are among the most fundamental problems in biology. In fact, many theoretical hypothesis on different type of speciation have been proposed. In particular, models of sympatric speciation leading to the formation of new species without geographical isolation, are based on the niche hypothesis: the diversification of the population is induced by the competition for a limited set of the available resources. On the other hand, neutral models of evolution have shown that stochastic forces are sufficient to generate coexistence of different species. In this work, we bring this dichotomy to the context of species formation, and we study how neutral and niche forces contribute to sympatric speciation in a model ecosystem.
In particular, we study the evolution of a population of individuals with asexual reproduction whose inherited characters or phenotypes are specified by both niche-based and neutral traits. We analyse the stationary state of the dynamics, and study the distribution of individuals in the whole space of possible phenotypes. We show, both by numerical simulations and analytics, that there is a non-trivial coupling between neutral and niche forces induced by stochastic effects in the evolution of the population that allows the formation of clusters (i.e., species) in the phenotypic space. Our framework can be generalised also to sexual reproduction or other type of population dynamics.
\end{abstract}

\maketitle

\section{Introduction}
One of the fundamental problems in theoretical biology is the search for key mechanisms
leading to the emergence of biodiversity \cite{Doebeli0,pigolotti2007,Aguiar,McKane1,DaSilvia}. The understanding of the underlying processes at the origin of species diversifications are also essential for the maintenance and conservation of natural ecosystems biodiversity \cite{Magurran,Rosenzweig} and to predict how species adapt to changing external conditions \cite{libroDoebeli}.

Roughly speaking, we can distinguish two main different mechanisms for the formation of species: allopatric and sympatric \cite{Doebeli0}\footnote{We do not discuss here other mechanisms such as peripatric and paripatric speciation, as they can be viewed as subforms of the other ones.}. The first case occurs when an initial population of individuals belonging to the same species splits into different isolated subsets, typically due to geographical reasons (e.g. a new river, changes in the landscape or the formation of a canyon), to an extent that prevents or interferes with genetic interchange. Instead, sympatric speciation leads to the formation of new species without geographical isolation and differentiation is due to ecological interactions between individuals in the population \cite{Doebeli0}. Such a diversification of the population is thought to be induced by the competition for a limited set of the available resources and corresponds to the case studied in this paper.

On the experimental side, the large time scales in play hinder an exhaustive sampling of the data.
Still, fossils data \cite{raup1982,newman1999,pigolotti2005}) and molecular phylogenetic techniques \cite{gomez2002,rull2008} can be used to infer speciation activities in ancient times and in different geographical locations; controlled experimental studies are rare and specific for some bacterial species (e.g. e. Coli \cite{sniegowski1997,barrick2009}). Therefore, any major theoretical contribution is key to describe these long-term experimental data and to understand the evolutionary processes driving the observed speciation rates \cite{bak1993,harvey1994,phylo1,phylo2,Stadler,houchmandzadeh2017}. 
In this context, stochastic models constitute a powerful tool \cite{Blythe} and have had great success in areas such as population genetics, as these models are able to predict the gene frequencies among a population \cite{lambert2008,lambert2010,ewens2012}.

One of the simplest and most widespread stochastic evolutionary models dates from the 1930s and was introduced independently by Fisher and Wright \cite{Blythe}. Such a simplicity derives from the so called neutral assumption, i.e. that all individuals/genes are equally likely to reproduce/be inherited \cite{Hubbell,Tilman2004,reviewNT}. Fluctuations still occur due to the random nature of demographic processes. By limiting the amount of available resources, the purely neutral dynamics converge to a state of monodominance where a single type of individual/gene remains in the population \cite{Blythe}. Only if we add the possibility of a small mutation probability, different types of individuals can coexist \cite{houchmandzadeh2017}. Therefore, in a neutral scenario, the emerging diversity is the results of two opposite forces (see Figure 1A). From one side, differentiation, i,e., stochastic mutations leading to a diffusion of the individuals in the parameters space. On the other hand we can observe cohesion, i.e. offsprings appear by birth close (in the parameter space) to its parent and thus have a similar genomes with small fluctuations - (phenomena known also as neutral clustering \cite{houchmandzadeh2008,houchmandzadeh2009}).


Certainly, the neutral hypothesis cannot be valid for all kind of individuals and species. However, such a simplification has been proven to be enough in order to explain the emergence of many statistical patterns in nature, as for instance in communities of species belonging  to the same trophic level, such as tropical forests or coral reefs \cite{amos1,amos2}.  In most of the neutral models, a mutation (or speciation) parameter is needed in order to reach coexistence of different species \cite{reviewNT}. Nevertheless, some models have also investigated the role of environmental noise \cite{kessler2015,hidalgo2017}, sexual reproduction and limited dispersal \cite{Aguiar} or competition \cite{martin2016} - finding in both cases a non trivial emergence of biodiversity. 

Niche theories \cite{Niche,chase2003ecological,Tilman2004,Nadav2014} constitute a completely different paradigm, putting the emphasis on competitive interactions. These theories state that formation of coexisting species is only possible through diversification for exploitation of the resources, minimizing competition among individuals \cite{pigolotti2010} (Figure 1B), and postulate that the number of coexisting species is equal to the number
of niches or ways to exploit the resources of the environment \cite{hardin1960} (this is what has been called the
niche dimension hypothesis \cite{harpole2007}). In other words, competition is considered by niche theories as one of the crucial drivers leading to the emergence of biodiversity.

\begin{figure}
 \centering\includegraphics[width=\columnwidth]{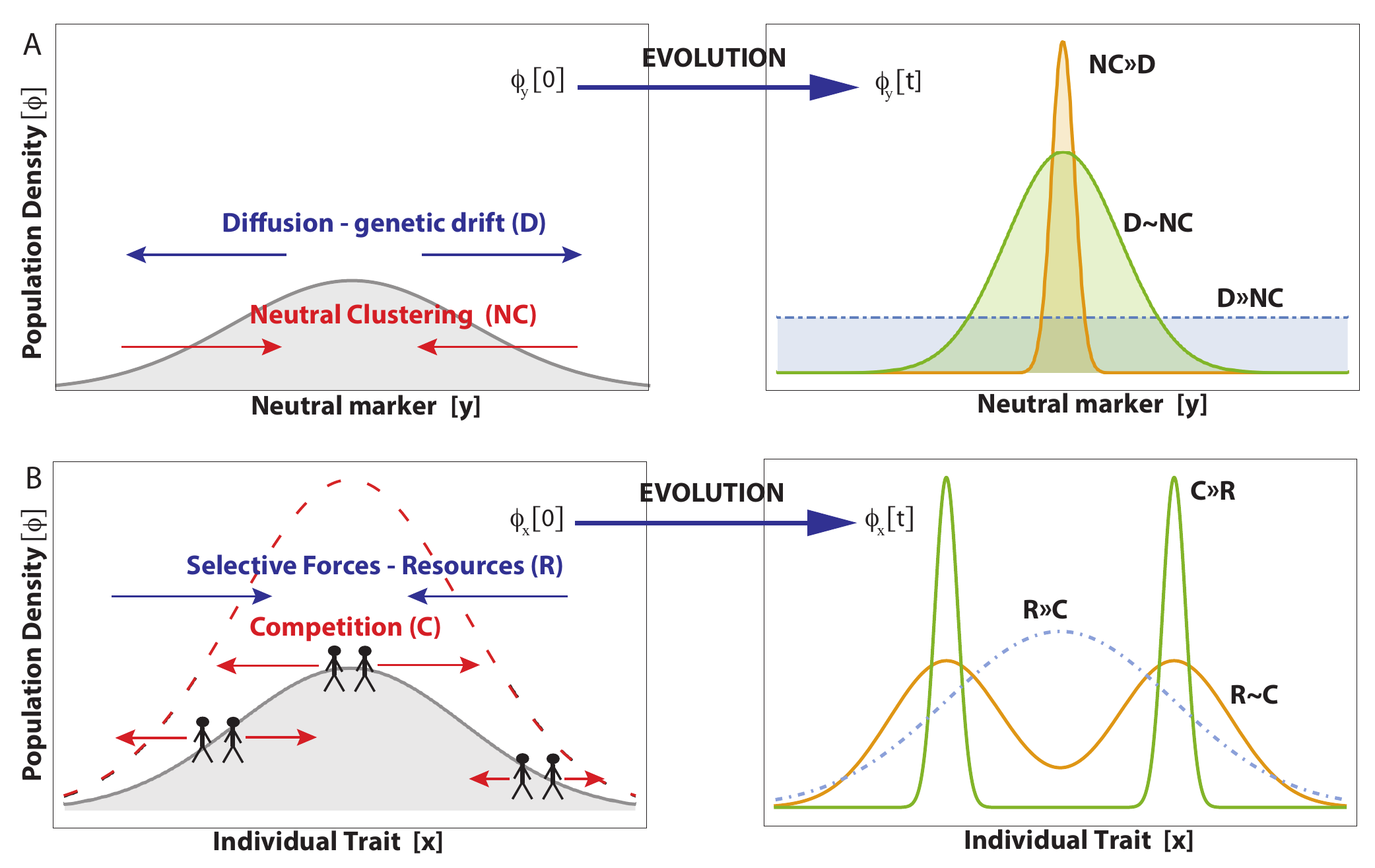}
 \caption{Sketch of neutral markers ($y$) and niche-based traits ($x$) evolution starting from the corresponding quasi-homogenous distribution $\Phi_y$ and $\Phi_x$ (in light grey). A) In the neutral case, the traits evolution is driven by genetic drift (D) and neutral clustering (NC): depending on the relative intensity of these two forces we can end up with a very peaked or very broad distribution.  B) On the niche case, traits undergo a selective force (R) pushing towards the region of higher resources (dash red line), and a repulsive force (C) among individuals so to minimize competition. Again the final distribution of traits will depend on the relative intensity between the two forces, and in this case in some regimes ($C\geq R$) sympatric speciation can be observed. In this work we want to study the evolution of a population of traits composed by both neutral and niche-based traits ($\Phi(x,y)$).}
 \label{fig:1}
\end{figure}
 
During the last decade, many works have demonstrated that the niche and the neutral paradigms
are not contradictory, but rather two complementary extreme views of what actually occurs in nature
\cite{leibold2006,chisholm2010niche,haegeman2011,fisher2014}.  In this paper, we bring this dichotomy to the context of species formation and make the following question:  Do new speciation mechanisms emerge when these forces are coupled together?

To answer this question, we investigate a set of models with the main ingredients of both niche and neutral theories.  In our simplest analysis, we study the evolution of a population of individuals whose characteristics are given by two sets of variables: the first set is niche-based and determines the way in which individuals exploit the (limited) resources,
while the second one is neutral and exclusively related to the inheritance process.
If only one set of these variables were present, one might expect clusterization of individuals in the space of traits due to cohesion or competition, respectively. Interestingly, we find that stochastic effects induce a non-trivial coupling between neutral and niche forces leading to correlations across different sets of traits, and clusters emerging in the 
niche-based space can be identified with clusters in the neutral space.

\section{The Model}\label{sec:1}
In our model, each individual $i$ in the population is represented by a pair of coordinates in a continuous phenotypic space with periodic boundary conditions, $(x_i, y_i)\in[L\times L]$ ($L$ represents a scale in the phenotypic space). Coordinate $x_i$ is a ``niche-based'' trait determining the way in which individuals exploit the resources, and $y_i$ is a ``neutral'' trait that has no impact on their fitness and simply acts as a (neutral) marker (see Fig. \ref{fig:2}). Resources are uniformly distributed and therefore no phenotypes are (a priori) fitter than other others. We have also analyzed the case in which individuals are represented by binary strings, rather than by two coordinates. The analysis becomes more complicated than for the continuous-space representation but our main conclusions remains the same.

In the dynamics, each individual reproduces at a constant rate $ b = 1 $. Offspring inherits both traits from their ancestor with a small, random mutation (see below for details). Competition for the resources is introduced through a variable death rate, $d(x)$, proportional to the number of individuals competing for the same resource, which only depends on the first trait:
\begin{equation}\label{c}
d(x_{i}) \; = \; \frac{1}{\Omega} \; \sum_{j \; \neq \; i} K_{c} (\mid x_{i} \; - \; x_{j} \mid \; ; \; w_{c}).
\end{equation}
$\Omega$ is the carrying capacity of the system and $K_c$ is a competition kernel that depends on the phenotypic distance between individuals (with the boundary conditions), that we take in terms of the Heaviside function $\Theta$:
\begin{equation}
K_{c} (z \; ; \; w_{c}) \; = \;\frac{ \Theta (w_{c} \; - \; z)}{2 w_{c}}.
\end{equation}
The range of competition is represented by $w_c$. In some situations, the choice of the kernel is crucial as different functional forms might lead to different outcomes \cite{libroDoebeli}. However, this is in general an analytical problem and does not present issues on individual-based implementations \cite{McKane1}. Our model is also suitable for investigating sexual reproduction (in the spirit of \cite{lafuerza}). For instance, one can take the birth rate as a function of the density of individuals with similar traits (assortative mating). In this case, the birth rate is not constant, but it grows with density of individuals with similar neutral traits (both $x$ and $y$). Our conclusions are robust to this extension.

Mutations are implemented as Gaussian, independent deviations on the coordinates $x$ and $y$ with zero mean and variance $\mu_x$ and $\mu_y$, respectively (considering periodic boundary conditions).  Different parameter values fix the balance between diffusion and cohesion for each trait. Numerical simulations are done using a Gillespie algorithm \cite{Gillespie} with $\mu_x$ and $\mu_y$ being the key control parameters. Generations are thus overlapping, and the number of individual in the population changes over time, fluctuating around the carrying capacity $\Omega$ after an initial transient time. A similar phenomenology can be obtained using the Wright-Fisher scheme with non-overlapping generations and a fixed population size, which provides a number of numerical advantages (see Appendix A) but entails a more difficult mathematical analysis.

\subsection{Mesoscopic description}
To gain some insight on the phenomenology, we can integrate a set of stochastic equations that describes the dynamics for large (but finite) population sizes, that we call the mesoscopic description.
This approach is useful as $i)$ it provides a more efficient way to integrate the dynamics for large populations, $ii)$ allows for simple mathematical treatment and $iii)$ highlights the ingredients that are key in the observed phenomenology. To derive these equations, the phenotypic space is discretized in small bins of size $\Delta\times\Delta$. Then one writes an equation for the number of individuals in $[x,x+\Delta]\times[y,y+\Delta]$ as $\phi(x,y,t)\Delta^2$, where $\phi(x,y,t)$ is the population density function at $(x,y)$ and time $t$. For large populations ($\Omega\gg1$), one can perform a system-size expansion of the dynamics \cite{VanKampen}. Finally, taking $\Delta\rightarrow0$, we find (see Appendix B):
\begin{multline}
 \partial_t \phi(x,y,t) = b \phi(x,y,t)\left(1-\frac{K_\mathrm{eff}(x,t)}{\Omega} \right) + D_x \partial_x^2 \phi(x,y,t)\\
+ D_y \partial_y^2 \phi(x,y,t) + \Omega^{-1/2} \sqrt{\phi(x,y,t)} \xi(x,y,t),
 \label{eq:phi}
\end{multline}
where $\xi(x,y,t)$ is a Gaussian white noise with zero mean and correlation function $\langle \xi(x,y,t)\xi(x',y',t') \rangle = \delta(x-x') \delta(y-y') \delta(t-t')$ and the effective competition kernel $K_\mathrm{eff}$ is defined as:
\begin{equation}
 K_\mathrm{eff}(x,t)=\int dx' dy' K(|x-x'|) \phi(x',y',t).
\end{equation}
Let us notice the presence of multiplicative demographic noise, which plays a fundamental role in eventually taking the system to local extinction, $\phi(x,y)=0$ \cite{alhammal2005}. The numerical integration of Eq. \eqref{eq:phi} can be performed re-introducing the discretization in $(x,y)$ with a certain resolution. 

Alternatively, we can study analytically the properties of Eq. (\ref{eq:phi}) in Fourier space \cite{pigolotti2007,DaSilvia}. In particular, we need to calculate the dispersion relation $\lambda(k_x,k_y)$, describing the stability of the trivial uniform solution for both traits $\phi(x,y)\sim \phi_0$. If $\lambda(k_x,k_y)$ takes positive values for any positive $k_x,k_y$, then the cohesion forces causes $\phi_0$ to be unstable, and clusters do emerge.

\section{Results}
\begin{figure}
 \centering\includegraphics[width=0.87\columnwidth]{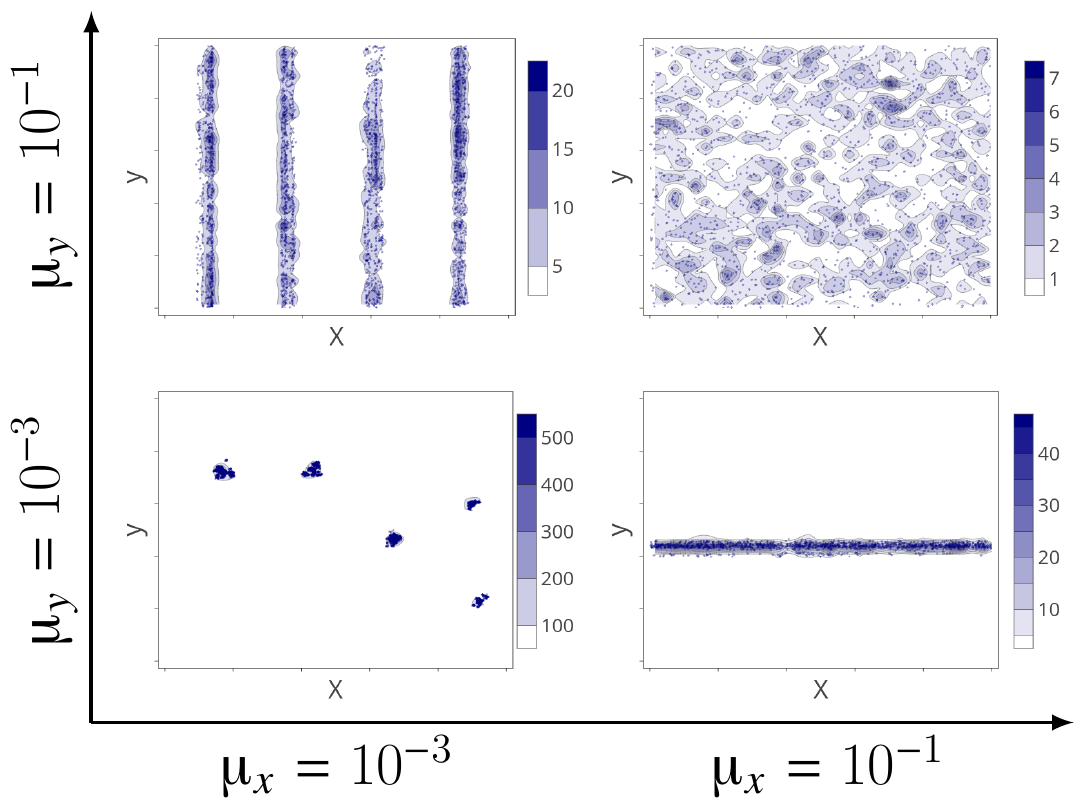} 
\caption{Results of numerical simulations run through the Gillespie Algorithm, with $\Omega = 10^3$ and $w_{c} = 0.2$. Both $x \in [0,1]$ and $y \in [0,1]$ with periodic boundary conditions.
Qualitatively, these results do not depend on the initial number of individuals (provided it is big enough to avoid eventual extinction) and of the initial population distribution in phenotypic space (e.g. random traits for the initial population or a common value for all individuals).}
 \label{fig:2}
\end{figure}

We first present the results obtained with the individual-based dynamics (run through the Gillespie Algorithm) for different values of the parameters (see Fig. \ref{fig:2}). For $\mu_x > \mu_{x}^{c}$, diffusion prevails and the niche-based trait $x$ is uniformly populated. For $\mu_x < \mu_{x}^{c}$, competitive forces split the population into isolated clusters. As expected, for fixed $\mu_x < \mu_{x}^{c}$ and $\Omega$, the spacing of the pattern observed along the niche-based trait $x$ depends only on $w_c$ and not on $\mu_y$, i.e., the niche interactions act only on $x$ and the neutral trait $y$ does not play any relevant role in the competition dynamics. These results can be perfectly framed within the dispersion relation as reported by \cite{pigolotti2007}.

On the other hand, the behaviour of the traits $y$ is far from being independent from the dynamics on $x$ and the phenomenology observed along the neutral trait is far less expected. The most interesting case corresponds to the situation in which $\mu_y < \mu_{y}^{c}$ and $\mu_x < \mu_{x}^{c}$ (i.e. when species form along $x$), where the neutral dynamics would predict the survival of a single cluster (because the system reaches an absorbing state). Instead, the population segregates into multiple isolated clusters in both the neutral and niche-based trait (see Fig. \ref{fig:2}). This phenomenon can be understood as a sort of neutral clustering \cite{houchmandzadeh2008,houchmandzadeh2009} within each of the species formed by competitive interactions, leading to different phenotypic pools of the neutral trait. In other words, speciation on the characters undergoing a competitive dynamics can lead to speciation on the characters undergoing a purely neutral dynamics. Finally, keeping small values of $\mu_y$ but taking $\mu_x > \mu_{x}^{c}$ (i.e. no speciation along $x$), neutral clustering emerges around a single value of $x$, as predicted by neutral clutering \cite{houchmandzadeh2008,houchmandzadeh2009}. When $\mu_y > \mu_{y}^{c}$, diffusion prevails and $y$ is populated uniformly independently on the value of $\mu_x$. 

To shed some light on these results, we start by analyzing the dynamics in very large populations, where demographic fluctuations can be neglected. 
To do that, we take the mean-field limit ($\Omega\rightarrow\infty$) of Eq. (\ref{eq:phi}), representing the infinite population limit where the stochastic term disappears, and we perturb the stationary solution $\phi_{\infty}= b/d$, i.e.  $\; \phi(x,y,t) = \phi_{\infty} + \varepsilon \; e^{i k_x x + i  k_y y + \lambda(k_x,k_y) t} \;$. With this procedure, we calculate the dispersion relation in Fourier space \cite{pigolotti2007,DaSilvia}, obtaining:

\begin{equation}\label{lambda_xy}
\lambda(k_x,k_y) \; = \; - \; d \; \phi_{\infty} \; 2 \pi \; \frac{sin(k_x\;T_c)}{k_x\;T_c} \frac{\delta(k_y)}{L_y}- \mu_x \; k_x ^2 - \mu_y \; k_y ^2
\end{equation}

\begin{figure}
 \centering\includegraphics[width=0.9\columnwidth]{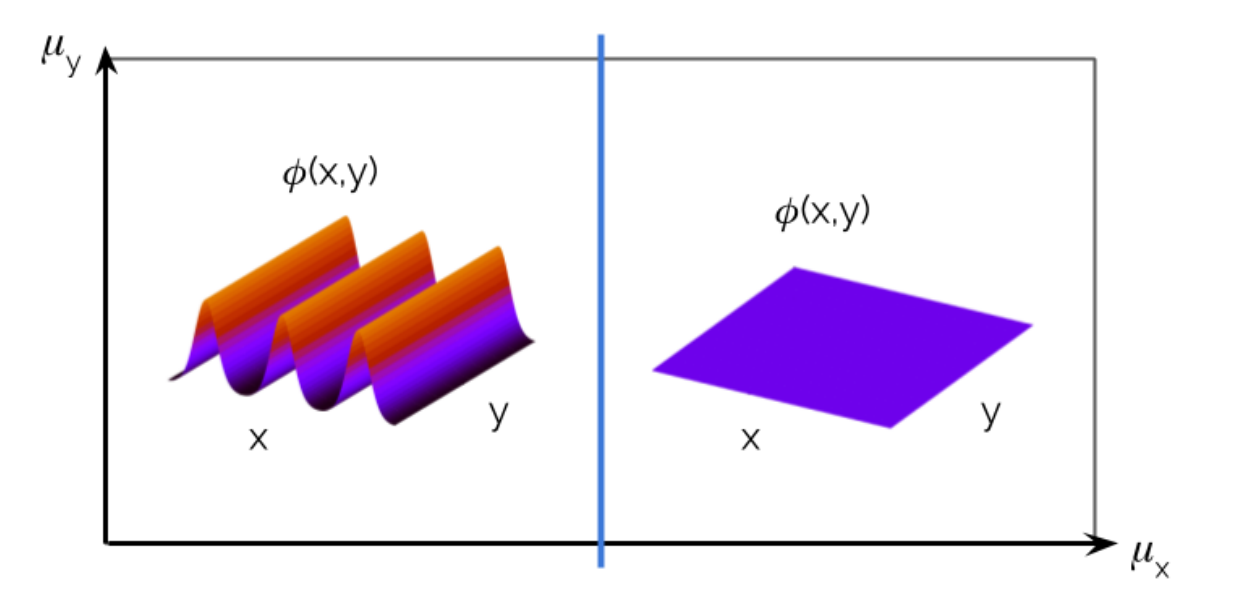}
 \caption{Schematic representation of the qualitative behavior observed when solving numerically the mean field limit of equation \ref{lambda_xy}. The mutation rate on the neutral trait $\mu_y$ plays a role only in fixing the time scale necessary to reach this steady configuration starting from any initial condition, but it does not affects the stationary solution.}
 \label{fig:MF}
\end{figure}

We find that in this case the mutation rate on the niche-base axis, $\mu_x$, becomes the critical parameter, whereas the mutation rate of the neutral-based trait, $\mu_y$, becomes irrelevant (see Fig. \ref{fig:MF}): the population density appears to be homogeneously distributed in phenotypic space for large values of $\mu_x$ (fluctuations around the homogeneous solution are stable, $\lambda(k_x,k_y)>0$); instead, fluctuations become unstable for small values of $\mu_x$ ($\lambda(k_x,k_y)>0$), and the population splits into different clusters along the niche-based trait, whereas it is still uniformly distributed along the neutral trait. If such neutral trait is neglected, this phenomenology is exactly the type of sympatric speciation due to competition for the resources reported by many authors in literature \cite{libroDoebeli, pigolotti2007,DaSilvia}. 
However, as we will show below, these results do not describe the phenomenology summarized in Fig. \ref{fig:2} and what actually occurs when implementing the full stochastic individual based dynamics, i.e. the mean field approximation, does not properly describe the system dynamics in this case. 

As a next step, we analyze the mesoscopic description given by the Karmer-Moyal expansion \cite{Gardiner2009,VanKampen} (Eq. (\ref{eq:phi})) for finite values of $\Omega$. Up to the first order, the expansion naively gives $D_{x}=2b\mu_{x}$ and $D_y=2b\mu_y$. However, numerical integrations of Eq. \eqref{eq:phi} with such values lead to radically different results than the one shown in Fig. \ref{fig:2}: the population density still does not clusterizes along the $y-$space, even for small values of $\mu_y$, and therefore neutral clusterization cannot be observed from a numerical integration of Eq. \eqref{eq:phi}. 

In the individual-based dynamics, clustering on the neutral axis emerges as a consequence of purely demographic fluctuations acting on the $x-$ trait, with highly populated regions that reproduce faster and low dense regions that are more likely to become extinct, reaching eventually the (local) absorbing state. However, the regularization power of diffusion in Eq. \eqref{eq:phi} plays in detrimental of this phenomenon, as any positive local density rapidly diffuses on the $y$-axis, thus recovering the system from local extinction to a positive (although small) density. In other words, `standard' diffusion in the equation avoids the emergence of local clusters.

\begin{figure}
 \centering\includegraphics[width=0.87\columnwidth]{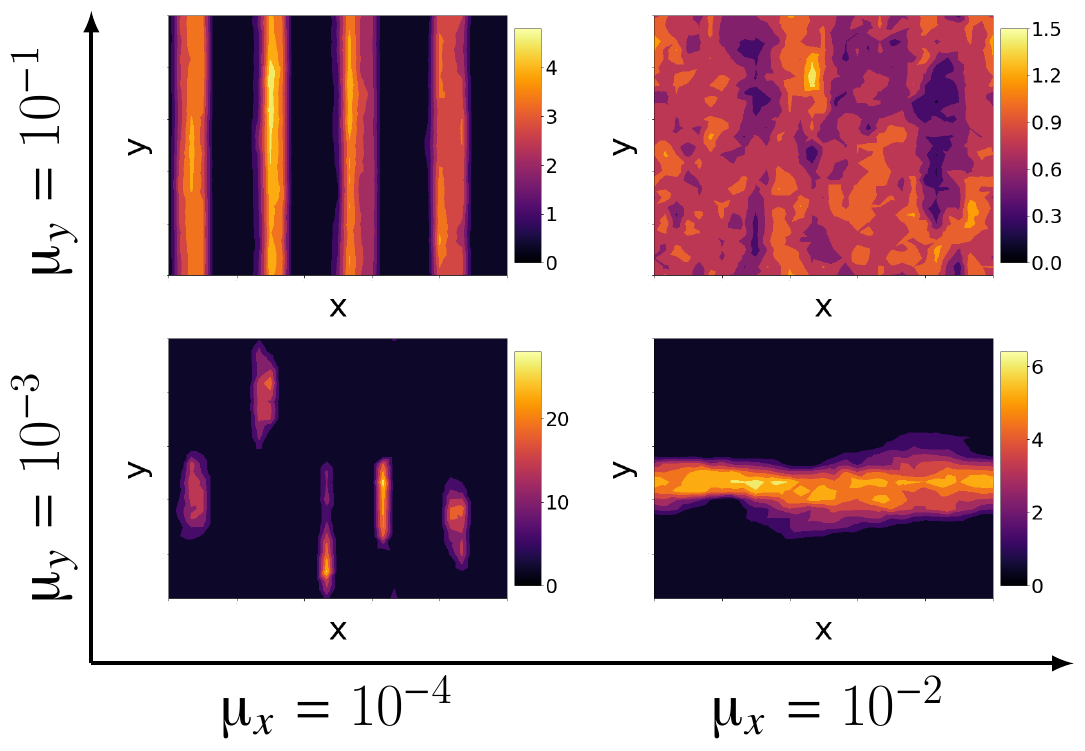}
\caption{Numerical integrations of the mesoscopic equation (Eq. \eqref{eq:phi}) with $\Omega = 1 $, $w_{c} = 0.15$, $b = 1$, $d=1$ and $\delta = 1. \;$ Both $x \in [0,1]$ and $y \in [0,1]$ with periodic boundary conditions.}
 \label{fig:3}
\end{figure}

As reported, the system-size expansion fails to capture the relevant ingredients of the underlying dynamics (other examples of the limitations of the system-size expansion can be found in \cite{diPatti2011}). To overcome this problem, one possibility would be to take next-to-leading terms that keep under control the undesired homogenizing effect. Alternatively, we could simply introduce in a heuristic manner a non-linearity in the diffusion term that takes into account that low population densities (where discrete effects become important and the mesoscopic description fails) are less likely to diffuse. A simple solution can be:
\begin{equation}
 D_{x} = 2 b \mu_{x} \frac{\phi(x,y)}{\delta+\phi(x,y)},
 \label{eq:diffusion}
\end{equation}
and similarly for $D_y$ and $\mu_y$. With this choice, diffusion becomes proportional to the population density for $\phi\ll \delta$ (thus limiting the diffusion of small populations), and constant if $\phi\gg \delta$, where $\delta$ is a new parameter controlling the crossover and should be taken proportional to the minimum population density in a discrete implementation, $\delta \sim \Omega^{-1}$. In this case the mesoscopic description and the individual based model give the same qualitatively behaviour, as shown in Fig. \ref{fig:3}. 

\section{Discussion and Conclusion}
In this work we have presented an individual based stochastic model, where each individual is described by both neutral and niche traits. 
The neutral traits represent those parts of the genome that simply diffuse in the genotype space through the effect of random mutations, while the niche ones are those traits that besides diffusing are also affected by selective forces (e.g. coding for the use of resources). Our interest was to understand the role of the two forces on speciation processes, i.e. the emergence of species in the system.

In pure neutral models, in the absence of mutations, fluctuations always cause the system to reach an absorbing state, i.e. by chance one species will start to predominate over the others, growing until it occupies all the system. Adding a mutation rate $\mu_x>0$ allows for coexistence of more species, where the number is proportional to the value of $\mu_x$ (continuously some species go extinct and new species enters the system).
On the other hand, in pure niche models competition is considered between similar individuals. This is generally ascribed to a limited resource availability, setting up an upper bound to the number of individuals that can get access to a particular type of nutrient. In this case the death rate is set to be proportional to the number of similar individuals and the effect of the competitive interactions is to separate the individuals in clusters, giving a maximization of the per-capita resources usage. The cluster formation is usually identified as the emergence of biodiversity, each cluster being a different species. 

In fact, the problem on how to define a species is not a trivial one. Indeed, it is still object of discussion between biologists. Many studies have proposed different definitions of species \cite{McKane0,McKaneLTT,houchmandzadeh2017}. Three type of species concepts are frequently employed in the literature: biological, ecological and genetic species. In populations undergoing sexual reproduction, a biological species is usually meant as a group of organisms that can mate only with other members of the same group. An eco-species is an ensemble of related organisms occupying a particular niche and having similar phenotypic traits (i.e., clusters along the $x$ axis). A genetic species is a group of organisms whose genomes are very similar to each other, and distinct from the individuals belonging to other clusters (e.g. clusters along the $y$ axis). In our model we consider both traits ($x,y$), and a species is given by the clustering of traits in both the $x$ and the $y$ coordinates, i.e. we consider eco-genetic species (Figures \ref{fig:2}-\ref{fig:3}).

We found that species emerge and coexist only in the range of low mutations rates for both the neutral and niche traits. This result highlights that dynamics induces a non-trivial coupling between the neutral and the competitive dynamics: clustering on the niche trait emerge as a consequence of a neutral dynamics acting on the $y-$ trait, i.e. local absorbing states induced by the neutral dynamics are reached in different regions of the $y$ axis, effectively clustering the genomes in both $x$ and $y$ coordinates. This coupling between the two (neutral and niche) forces is not caught by simple mean field nor Kramer-Moyal Gaussian approximations. In fact, the standard diffusion term arising in these cases does not allow to reach local extinction, forbidding the emergence of clusters at the individual level in both the $x$ and $y$ traits. In other words, the behaviour of the system in the mean field approximation is a simple and trivial combination of the features given individually by competition and neutral forces, without showing coupling effects of any sort between the two parts of the genotype.

In summary, we propose a general framework to study the emergence of biodiversity in biological systems by sympatric speciation mechanisms. Our models combine both neutral and niche-based features. We have observed that these two forces are intertwined in their contribution to speciation. In particular, niche forces promote the emergence of clustering on traits driven by neutral dynamics. 
The proposed framework can be generalized to consider other important biological mechanisms that may play an important role in the speciation processes such as horizontal gene transfer, optimization of the use of resources and the effect of environmental fluctuations. We plan to investigate these aspects in future works. 

\begin{acknowledgments}
  The authors would like to thank Davide Biraghi for preliminary work on this project. S.S acknowledges the University of Padova for funding (STARS  grant 2018).
\end{acknowledgments}

\section{Appendix A: Write-Fisher-explicit resource model}
\begin{figure}[h!]
 \centering\includegraphics[width=0.87\columnwidth]{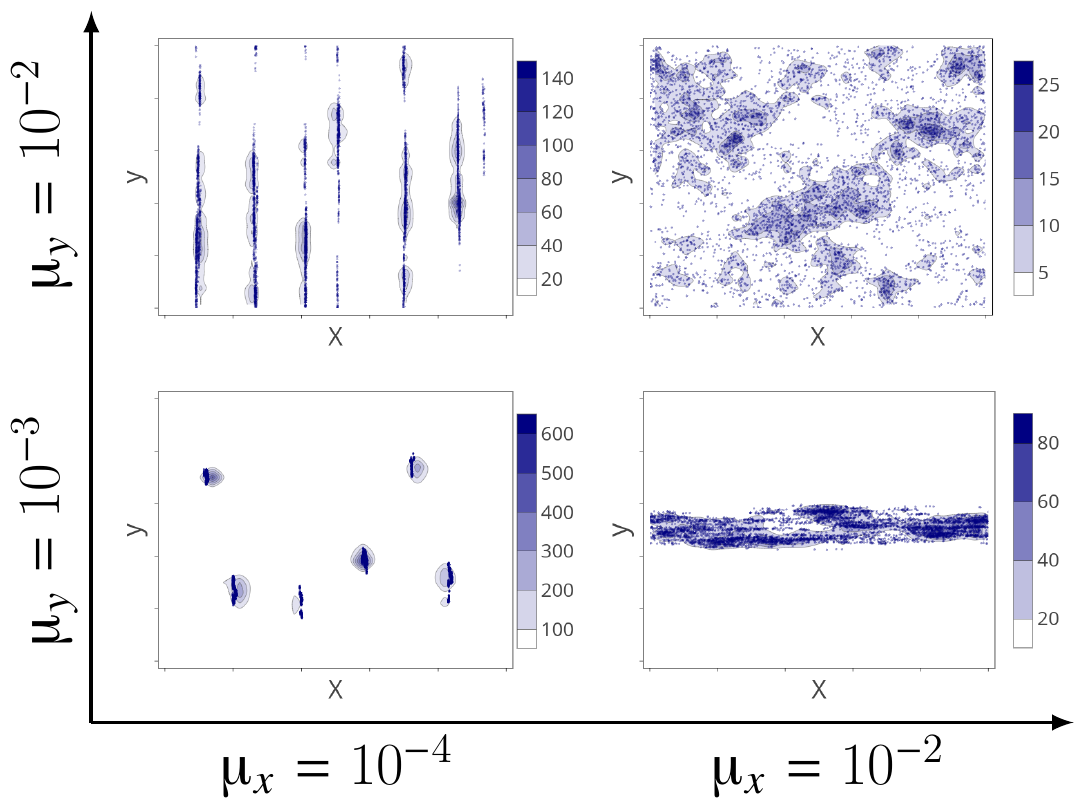}
\caption{Results of numerical simulations run in Wright Fisher simulation scheme, with non-overlapping generations and explicit representation of resources. The population size is $N = 5 \cdot 10^3$ and the competitive interaction range is $w_{c} = 0.2$. Both $x \in [0,1]$ and $y \in [0,1]$ with periodic boundary conditions using a discretization $ \Delta x = \Delta y = 2 \cdot 10^{-3} $; the simulations are run for $N_{gen} = 5 \cdot 10^3$ generations and with $N_{trials} = 25 * 10^4$. Our results confirm that this simulation scheme describes the process under study as  well the Gilliespie simulations and the mesoscopic phenomenological Equation (eq. \ref{eq:phi}).}
 \label{fig:4}
\end{figure}

We studied the same dynamics as described by eq. \ref{eq:phi} also by means of numerical simulations of a Wright-Fisher-like model, with fixed population size $N$ and non-overlapping generations \cite{Blythe}. Within this dynamics, we modelled the competition for the resources in an explicit way, by representing the resources as a second population of fixed finite size, made of entities with the same structural properties of the individuals in the population (that is, they  take the form of binary strings or of single real variables depending on which of these representations is chosen). In our simulations, we can divide the evolution from a generation to the next one in two steps: selection and replication. Each item being tested at the step of selection will have to find a resource compatible with its characteristics, that is, similar too it up to a given maximum discrepancy threshold. Limitation of resources is modeled as it follows: for every possible resource type there is an equal limited set of copies; if an individuals uses a resource when it is selected, that resource will not be available for the others individuals in that generation.\\
Each simulation proceeds as follows. First, given a population and a resource pool, we extract with uniform probability one random individual and one resource item; if they are compatible with each other, they are removed from their own starting sets, and the individual is added to a pool of selected ones. If they are not compatible, they are reinserted in their respective living communities. This procedure is repeated until an exit condition becomes satisfied, being it the fact that all the individuals in the community have been selected or that the procedure has reached a fixed number of iterations $N_{trials}$. Once that a selected pool of individuals has been identified, all the remaining ones are discarded, and a new generation is built up by picking at random from the selected entities with uniform probability; the selected items are replicated with mutations, aimed to mimic the genetic drift, until the new population reaches the same size $N$. The whole procedure is repeated for a fixed number of generation $N_{gen}$. Our simulations (Figure \ref{fig:4}) exhibit the same qualitative behavior observed in the Gillespie simulations (Figure \ref{fig:2}) and the numerical integrations of eq. \ref{eq:phi} (Figure \ref{fig:3}). 
This simulation scheme has the advantage of being computationally faster than Gillespie simulations, since it does not require to calculate explicitly pair interactions among all individuals. However, it entails a more difficult mathematical analysis.

\section{Appendix B: Phenomenological description of the model}
Equation \eqref{eq:phi} can be derived explicitly starting from the definition of the microscopic processes occurring in the dynamics \cite{McKane1, Blythe}. Here we show how to obtain an equivalent equation for a slightly simplified case. We focus on a population whose individual's phenotypes are represented by single real valued variables $x$, and consider the evolution under the effect of asexual reproduction, competition for resources, mutations and demographic stochasticity. As before, $\phi(x,t)$ is the population density function of individuals $x$ at time $t$. It is useful to write it as 
\begin{equation}\label{eq:density}
\phi(x,t) = \cfrac{1}{\Omega} \sum_{i=1}^{N(t)}\delta(x-x_i)
\end{equation}
 where $N(t)$ is the number of individuals present at time $t$ and $\Omega$ is the characteristic size of the system. In this view, the birth/death of an individual with phenotype $x'$ corresponds to adding/subtracting to the density \eqref{eq:density} a Dirac delta centred on $x'$. The microscopic transition rates defining the probability of birth and death of an individual $x'$ can be written as
 
\begin{equation}\label{eq:rates}
\begin{gathered}
   W \left ( \phi(x) \xrightarrow{} \phi(x)+\cfrac{1}{\Omega} \; \delta(x-x') \right ) = b \; \Omega \; \phi(x') \; \mathcal{M}(x-x') \\
  W \left( \phi(x) \xrightarrow{} \phi(x)-\cfrac{1}{\Omega} \; \delta(x-x') \right)  = d \; \Omega  \; \phi(x') \; \delta(x-x') \; \times\; \\
  \times \int dz \; \phi(z) \; \mathcal{K}_c(z-x')
\end{gathered}
\end{equation}
The first rate states that the probability of birth of an individual $x'$ is proportional to the density $\phi(x)$, modulated by a mutation kernel $\mathcal{M}(x-x')$ centred on $x'$. We consider here mutation as Gaussian noise added to the parents phenotype $x'$, so that the mutation kernel takes the form:
\begin{equation}
  \mathcal{M}(x-y) = \cfrac{1}{\sqrt{2 \pi (2 D)}} \; \textit{e}^{- \frac{(x-y)^2}{2 \pi  (2 D)}}
\end{equation}
At the same time, the second rate implies that the probability of death of an individual $x'$ is proportional to the product of the density $\phi(x')$ and the number of individuals similar to $x'$: $\int dz \; \phi(z) \; \mathcal{K}_c(z-x')$, where $\mathcal{K}_c(z-x')$ is the competition kernel. We can now write a Master Equation (ME) for the evolution of the probability density $P(\phi(x),t)$:
\begin{equation}\label{eq:ME}
\begin{gathered}
 \partial_t ^{} P(\phi(x), t) \;  = \\  \int dx'   W ( \phi(x)+\tfrac{1}{\Omega}   \delta(x-x')  \xrightarrow{} \phi(x) ) \\    P(\phi(x)+\tfrac{1}{\Omega}   \delta(x-x') , t) \\ 
  + \int dx'   W ( \phi(x)-\tfrac{1}{\Omega}   \delta(x-x')  \xrightarrow{} \phi(x) )  \\  P(\phi(x)-\tfrac{1}{\Omega}   \delta(x-x') , t) \\
  -  \int dx'   W ( \phi(x) \xrightarrow{} \phi(x)+\tfrac{1}{\Omega}   \delta(x-x') )    P(\phi(x), t)  \\
  -  \int dx'   W ( \phi(x) \xrightarrow{} \phi(x)-\tfrac{1}{\Omega}   \delta(x-x') )    P(\phi(x), t).
\end{gathered}
\end{equation}
Introducing the step operators $\Delta^+_{x'}$ and $\Delta^+_{x'}$, whose action on a generic functional $F[\phi(x)]$ is defined as:
\begin{equation}\label{eq:steps}
   \Delta_{x'}^{\pm} f [\phi(x) ] = f \left[ \phi (x) \pm \cfrac{1}{\Omega} \; \delta(x-{x'}) \right] 
\end{equation}
the ME takes the form:
\begin{equation}\label{eq:MEsteps}
\begin{gathered}
\partial_t ^{} P(\phi(x), t) = \\
 \int d x'   ( \Delta_{x'}^- -1) \left( W  ( \phi(x) \xrightarrow{} \phi(x)+\tfrac{1}{\Omega}   \delta(x- x') )    P(\phi(x), t) \right ) \\
 +  \int d x'   ( \Delta_{x'}^+ - 1) \left( W  ( \phi(x) \xrightarrow{} \phi(x)-\tfrac{1}{\Omega}   \delta(x- x') )    P(\phi(x), t) \right ) 
\end{gathered}
\end{equation}
We can now perform a system size expansion on the step operators, which can be viewed as a functional variant of the Kramers-Moyal expansion, valid in the large population regime:
\begin{equation}\label{eq:expansion}
   \Delta_{x'}^{\pm} = 1 \pm \cfrac{1}{\Omega} \; \cfrac{\delta}{\delta \phi(x')} + \cfrac{1}{2 \Omega^2} \; \cfrac{\delta ^ 2}{\delta \phi ^ 2 (x')} \; + O \left ( \cfrac{1}{\Omega ^ {3}} \right ),
\end{equation}
where $\delta / \delta(x)$ denotes a functional differentiation. By inserting into Eq.\eqref{eq:MEsteps} the expansion given by Eq.\eqref{eq:expansion} truncated at the second order, we obtain to the Fokker-Plank equation (FPE):
\begin{equation}\label{eq:fokkerplank}
\begin{gathered}
\partial_t ^{} P(\phi(x), t)   \approx \\
- \int d x'   \cfrac{\delta}{\delta \phi( x')} 
 \bigg[  \cfrac{1}{\Omega}  P(\phi(x),t) \Big( W  ( \phi(x) \xrightarrow{} \phi(x)+\tfrac{1}{\Omega}   \delta(x- x')) - \\ W  ( \phi(x) \xrightarrow{} \phi(x)-\tfrac{1}{\Omega}   \delta(x- x')) \Big) \bigg] \\
 +   \cfrac{1}{2} \int d x'  \cfrac{\delta ^2}{\delta \phi^2( x')} 
 \bigg[ \cfrac{1}{\Omega^2}  P(\phi(x),t) \Big( W  ( \phi(x) \xrightarrow{} \phi(x)+\tfrac{1}{\Omega}   \delta(x- x')) +  \\ W  ( \phi(x) \xrightarrow{} \phi(x)-\tfrac{1}{\Omega}   \delta(x- x')) \Big) \bigg]
\end{gathered}
\end{equation}
which in turn can be mapped into the Langevin equation
\begin{equation}\label{eq:langevin}
\begin{gathered}
\partial_t ^{} \phi(x, t) \;  = \int d x' \; \cfrac{1}{\Omega} \; W  ( \phi(x) \xrightarrow{} \phi(x)+\tfrac{1}{\Omega} \; \delta(x- x') ) \; - \\
 \int d x' \; \cfrac{1}{\Omega}\;  W  ( \phi(x) \xrightarrow{} \phi(x)-\tfrac{1}{\Omega} \; \delta(x- x') ) \; + \\
  \int d x' \; \sqrt{ \cfrac{1}{\Omega^2} \; \Big ( W  ( \phi(x) \xrightarrow{} \phi(x)+\tfrac{1}{\Omega} \; \delta(x- x') ) \;+ \;\; } \\ \overline{ \; W  ( \phi(x) \xrightarrow{} \phi(x)-\tfrac{1}{\Omega} \; \delta(x- x') ) \Big ) } \cdot \xi (x, t).
\end{gathered}
\end{equation}
$\xi (x, t)$ is a Gaussian zero-mean white noise with correlation function $\left \langle \xi (x, t)\xi (x', t') \right \rangle =\delta(x-x')\delta(t-t')$.
Using the explicit form of the transition rates \eqref{eq:rates}, Eq.\eqref{eq:langevin} becomes:
\begin{equation}\label{eq1}
\begin{gathered}
\partial_t ^{} \phi(x, t) \;  = \; b \int d x' \; \phi( x') \; \mathcal{M} (x- x') \\- \; d \; \phi(x,t) \int d x' \; \phi( x', t) \; \mathcal{K}_c( x'-x) \\
 + \; \cfrac{\sqrt{\phi(x,t)}}{\sqrt{\Omega}} \; \sqrt{\left ( b + d \int d x'' \; \phi( x'', t) \; \mathcal{K}_c( x''-x) \right ) } \cdot \xi (x, t).
\end{gathered}
\end{equation}
For small enough values of the mutation parameter $D$ we can approximate the expression of the birth term as
\begin{equation}
\begin{gathered}
b \int dy \; \phi(y) \; \mathcal{M} (x-y) =  b \int d x' \Big[ \phi(x) +  \\ ( x'-x) {\phi}'(x) + \tfrac{1}{2} ( x'-x)^2   {\phi(x)}''   +   ...    \Big ]  \cfrac{e^{- \frac{(x- x')^2}{2 \pi (2D)}}}{\sqrt{2 \pi (2D)}} \\
\approx   b\; \phi(x)   +  b\;  D   \cfrac{\partial^2 \phi(x)}{\partial x^2}.
\end{gathered}
\end{equation}\\

Finally, approximating the demographic-noise term, we obtain to the equation:
\begin{equation}\label{eq:1Dfinal}
\begin{gathered}
\cfrac {\partial \phi(x, t)}{\partial t} \; = \; b \; \phi(x,t) - d \; \phi(x,t) \int dx' \; \phi(x', t) \; \mathcal{K}_c(x'-x) \; \\ + \; D \; \cfrac{\partial^2 \phi(x)}{\partial x^2}\;
 + \cfrac{1}{\sqrt{\Omega}} \; \sqrt{\phi(x,t)} \cdot \xi (x, t).
\end{gathered}
\end{equation}\\
It is now possible to see how Eq. \eqref{eq:phi} is just an extension of Eq. \eqref{eq:1Dfinal}, but where the phenotypes of any individual is as set of two real valued variables $(x,y)$ instead of only one $(x)$, and where competition is determined by the only trait set $x$, while mutation, asexual reproduction and demographic noise affects both characters.


\bibliography{neutral-niche}

\begin{thebibliography}{54}%
\makeatletter
\providecommand \@ifxundefined [1]{%
 \@ifx{#1\undefined}
}%
\providecommand \@ifnum [1]{%
 \ifnum #1\expandafter \@firstoftwo
 \else \expandafter \@secondoftwo
 \fi
}%
\providecommand \@ifx [1]{%
 \ifx #1\expandafter \@firstoftwo
 \else \expandafter \@secondoftwo
 \fi
}%
\providecommand \natexlab [1]{#1}%
\providecommand \enquote  [1]{``#1''}%
\providecommand \bibnamefont  [1]{#1}%
\providecommand \bibfnamefont [1]{#1}%
\providecommand \citenamefont [1]{#1}%
\providecommand \href@noop [0]{\@secondoftwo}%
\providecommand \href [0]{\begingroup \@sanitize@url \@href}%
\providecommand \@href[1]{\@@startlink{#1}\@@href}%
\providecommand \@@href[1]{\endgroup#1\@@endlink}%
\providecommand \@sanitize@url [0]{\catcode `\\12\catcode `\$12\catcode
  `\&12\catcode `\#12\catcode `\^12\catcode `\_12\catcode `\%12\relax}%
\providecommand \@@startlink[1]{}%
\providecommand \@@endlink[0]{}%
\providecommand \url  [0]{\begingroup\@sanitize@url \@url }%
\providecommand \@url [1]{\endgroup\@href {#1}{\urlprefix }}%
\providecommand \urlprefix  [0]{URL }%
\providecommand \Eprint [0]{\href }%
\providecommand \doibase [0]{http://dx.doi.org/}%
\providecommand \selectlanguage [0]{\@gobble}%
\providecommand \bibinfo  [0]{\@secondoftwo}%
\providecommand \bibfield  [0]{\@secondoftwo}%
\providecommand \translation [1]{[#1]}%
\providecommand \BibitemOpen [0]{}%
\providecommand \bibitemStop [0]{}%
\providecommand \bibitemNoStop [0]{.\EOS\space}%
\providecommand \EOS [0]{\spacefactor3000\relax}%
\providecommand \BibitemShut  [1]{\csname bibitem#1\endcsname}%
\let\auto@bib@innerbib\@empty
\bibitem [{\citenamefont {Dieckmann}\ and\ \citenamefont
  {Doebeli}(1999)}]{Doebeli0}%
  \BibitemOpen
  \bibfield  {author} {\bibinfo {author} {\bibfnamefont {U.}~\bibnamefont
  {Dieckmann}}\ and\ \bibinfo {author} {\bibfnamefont {M.}~\bibnamefont
  {Doebeli}},\ }\href@noop {} {\bibfield  {journal} {\bibinfo  {journal}
  {Nature}\ }\textbf {\bibinfo {volume} {400}} (\bibinfo {year}
  {1999})}\BibitemShut {NoStop}%
\bibitem [{\citenamefont {Pigolotti}\ \emph {et~al.}(2007)\citenamefont
  {Pigolotti}, \citenamefont {Lopez},\ and\ \citenamefont
  {Hernandez-Garcia}}]{pigolotti2007}%
  \BibitemOpen
  \bibfield  {author} {\bibinfo {author} {\bibfnamefont {S.}~\bibnamefont
  {Pigolotti}}, \bibinfo {author} {\bibfnamefont {C.}~\bibnamefont {Lopez}}, \
  and\ \bibinfo {author} {\bibfnamefont {E.}~\bibnamefont {Hernandez-Garcia}},\
  }\href@noop {} {\bibfield  {journal} {\bibinfo  {journal} {Phys. Rev. Lett}\
  }\textbf {\bibinfo {volume} {98}} (\bibinfo {year} {2007})}\BibitemShut
  {NoStop}%
\bibitem [{\citenamefont {de~Aguiar}\ \emph {et~al.}(2009)\citenamefont
  {de~Aguiar}, \citenamefont {Baranger}, \citenamefont {Baptestini},
  \citenamefont {Kaufman},\ and\ \citenamefont {Bar-Yam}}]{Aguiar}%
  \BibitemOpen
  \bibfield  {author} {\bibinfo {author} {\bibfnamefont {M.~A.~M.}\
  \bibnamefont {de~Aguiar}}, \bibinfo {author} {\bibfnamefont {M.}~\bibnamefont
  {Baranger}}, \bibinfo {author} {\bibfnamefont {E.~M.}\ \bibnamefont
  {Baptestini}}, \bibinfo {author} {\bibfnamefont {L.}~\bibnamefont {Kaufman}},
  \ and\ \bibinfo {author} {\bibfnamefont {Y.}~\bibnamefont {Bar-Yam}},\
  }\href@noop {} {\bibfield  {journal} {\bibinfo  {journal} {Nature}\ }\textbf
  {\bibinfo {volume} {460}} (\bibinfo {year} {2009})}\BibitemShut {NoStop}%
\bibitem [{\citenamefont {Rogers}\ and\ \citenamefont
  {Rossberg}(2012{\natexlab{a}})}]{McKane1}%
  \BibitemOpen
  \bibfield  {author} {\bibinfo {author} {\bibfnamefont {A.~J.}\ \bibnamefont
  {Rogers}, \bibfnamefont {T.~McKane}}\ and\ \bibinfo {author} {\bibfnamefont
  {A.~G.}\ \bibnamefont {Rossberg}},\ }\href@noop {} {\bibfield  {journal}
  {\bibinfo  {journal} {EPL (Europhysics Letters)}\ }\textbf {\bibinfo {volume}
  {97}} (\bibinfo {year} {2012}{\natexlab{a}})}\BibitemShut {NoStop}%
\bibitem [{\citenamefont {da~Silva}\ \emph {et~al.}(2014)\citenamefont
  {da~Silva}, \citenamefont {Colombo},\ and\ \citenamefont
  {Anteneodo}}]{DaSilvia}%
  \BibitemOpen
  \bibfield  {author} {\bibinfo {author} {\bibfnamefont {L.~A.}\ \bibnamefont
  {da~Silva}}, \bibinfo {author} {\bibfnamefont {E.~H.}\ \bibnamefont
  {Colombo}}, \ and\ \bibinfo {author} {\bibfnamefont {C.}~\bibnamefont
  {Anteneodo}},\ }\href@noop {} {\bibfield  {journal} {\bibinfo  {journal}
  {Phys. Rev. E}\ }\textbf {\bibinfo {volume} {90}} (\bibinfo {year}
  {2014})}\BibitemShut {NoStop}%
\bibitem [{\citenamefont {Magurran}(2004)}]{Magurran}%
  \BibitemOpen
  \bibfield  {author} {\bibinfo {author} {\bibfnamefont {A.}~\bibnamefont
  {Magurran}},\ }\href@noop {} {\emph {\bibinfo {title} {Measuring biology
  diversity}}}\ (\bibinfo  {publisher} {Blackwell Publishing, Oxford},\
  \bibinfo {year} {2004})\BibitemShut {NoStop}%
\bibitem [{\citenamefont {Rosenzweig}(1995)}]{Rosenzweig}%
  \BibitemOpen
  \bibfield  {author} {\bibinfo {author} {\bibfnamefont {M.~L.}\ \bibnamefont
  {Rosenzweig}},\ }\href@noop {} {\emph {\bibinfo {title} {Species diversity in
  space and time}}}\ (\bibinfo  {publisher} {Cambridge University Press},\
  \bibinfo {year} {1995})\BibitemShut {NoStop}%
\bibitem [{\citenamefont {Doebeli}(2011)}]{libroDoebeli}%
  \BibitemOpen
  \bibfield  {author} {\bibinfo {author} {\bibfnamefont {M.}~\bibnamefont
  {Doebeli}},\ }\href@noop {} {\emph {\bibinfo {title} {Adaptive
  diversification}}}\ (\bibinfo  {publisher} {Princeton University Press},\
  \bibinfo {year} {2011})\BibitemShut {NoStop}%
\bibitem [{Note1()}]{Note1}%
  \BibitemOpen
  \bibinfo {note} {We do not discuss here other mechanisms such as peripatric
  and paripatric speciation, as they can be viewed as subforms of the other
  ones.}\BibitemShut {Stop}%
\bibitem [{\citenamefont {Raup}\ and\ \citenamefont
  {Sepkoski}(1982)}]{raup1982}%
  \BibitemOpen
  \bibfield  {author} {\bibinfo {author} {\bibfnamefont {D.~M.}\ \bibnamefont
  {Raup}}\ and\ \bibinfo {author} {\bibfnamefont {J.~J.}\ \bibnamefont
  {Sepkoski}},\ }\href@noop {} {\bibfield  {journal} {\bibinfo  {journal}
  {Science}\ }\textbf {\bibinfo {volume} {215}},\ \bibinfo {pages} {1501}
  (\bibinfo {year} {1982})}\BibitemShut {NoStop}%
\bibitem [{\citenamefont {Newman}\ and\ \citenamefont
  {Sibani}(1999)}]{newman1999}%
  \BibitemOpen
  \bibfield  {author} {\bibinfo {author} {\bibfnamefont {M.}~\bibnamefont
  {Newman}}\ and\ \bibinfo {author} {\bibfnamefont {P.}~\bibnamefont
  {Sibani}},\ }\href@noop {} {\bibfield  {journal} {\bibinfo  {journal}
  {Proceedings of the Royal Society of London B: Biological Sciences}\ }\textbf
  {\bibinfo {volume} {266}},\ \bibinfo {pages} {1593} (\bibinfo {year}
  {1999})}\BibitemShut {NoStop}%
\bibitem [{\citenamefont {Pigolotti}\ \emph {et~al.}(2005)\citenamefont
  {Pigolotti}, \citenamefont {Flammini}, \citenamefont {Marsili},\ and\
  \citenamefont {Maritan}}]{pigolotti2005}%
  \BibitemOpen
  \bibfield  {author} {\bibinfo {author} {\bibfnamefont {S.}~\bibnamefont
  {Pigolotti}}, \bibinfo {author} {\bibfnamefont {A.}~\bibnamefont {Flammini}},
  \bibinfo {author} {\bibfnamefont {M.}~\bibnamefont {Marsili}}, \ and\
  \bibinfo {author} {\bibfnamefont {A.}~\bibnamefont {Maritan}},\ }\href@noop
  {} {\bibfield  {journal} {\bibinfo  {journal} {Proceedings of the National
  Academy of Sciences}\ }\textbf {\bibinfo {volume} {102}},\ \bibinfo {pages}
  {15747} (\bibinfo {year} {2005})}\BibitemShut {NoStop}%
\bibitem [{\citenamefont {G{\'o}mez}\ \emph {et~al.}(2002)\citenamefont
  {G{\'o}mez}, \citenamefont {Serra}, \citenamefont {Carvalho},\ and\
  \citenamefont {Lunt}}]{gomez2002}%
  \BibitemOpen
  \bibfield  {author} {\bibinfo {author} {\bibfnamefont {A.}~\bibnamefont
  {G{\'o}mez}}, \bibinfo {author} {\bibfnamefont {M.}~\bibnamefont {Serra}},
  \bibinfo {author} {\bibfnamefont {G.~R.}\ \bibnamefont {Carvalho}}, \ and\
  \bibinfo {author} {\bibfnamefont {D.~H.}\ \bibnamefont {Lunt}},\ }\href@noop
  {} {\bibfield  {journal} {\bibinfo  {journal} {Evolution}\ }\textbf {\bibinfo
  {volume} {56}},\ \bibinfo {pages} {1431} (\bibinfo {year}
  {2002})}\BibitemShut {NoStop}%
\bibitem [{\citenamefont {Rull}(2008)}]{rull2008}%
  \BibitemOpen
  \bibfield  {author} {\bibinfo {author} {\bibfnamefont {V.}~\bibnamefont
  {Rull}},\ }\href@noop {} {\bibfield  {journal} {\bibinfo  {journal}
  {Molecular Ecology}\ }\textbf {\bibinfo {volume} {17}},\ \bibinfo {pages}
  {2722} (\bibinfo {year} {2008})}\BibitemShut {NoStop}%
\bibitem [{\citenamefont {Sniegowski}\ \emph {et~al.}(1997)\citenamefont
  {Sniegowski}, \citenamefont {Gerrish},\ and\ \citenamefont
  {Lenski}}]{sniegowski1997}%
  \BibitemOpen
  \bibfield  {author} {\bibinfo {author} {\bibfnamefont {P.~D.}\ \bibnamefont
  {Sniegowski}}, \bibinfo {author} {\bibfnamefont {P.~J.}\ \bibnamefont
  {Gerrish}}, \ and\ \bibinfo {author} {\bibfnamefont {R.~E.}\ \bibnamefont
  {Lenski}},\ }\href@noop {} {\bibfield  {journal} {\bibinfo  {journal}
  {Nature}\ }\textbf {\bibinfo {volume} {387}},\ \bibinfo {pages} {703}
  (\bibinfo {year} {1997})}\BibitemShut {NoStop}%
\bibitem [{\citenamefont {Barrick}\ \emph {et~al.}(2009)\citenamefont
  {Barrick}, \citenamefont {Yu}, \citenamefont {Yoon}, \citenamefont {Jeong},
  \citenamefont {Oh}, \citenamefont {Schneider}, \citenamefont {Lenski},\ and\
  \citenamefont {Kim}}]{barrick2009}%
  \BibitemOpen
  \bibfield  {author} {\bibinfo {author} {\bibfnamefont {J.~E.}\ \bibnamefont
  {Barrick}}, \bibinfo {author} {\bibfnamefont {D.~S.}\ \bibnamefont {Yu}},
  \bibinfo {author} {\bibfnamefont {S.~H.}\ \bibnamefont {Yoon}}, \bibinfo
  {author} {\bibfnamefont {H.}~\bibnamefont {Jeong}}, \bibinfo {author}
  {\bibfnamefont {T.~K.}\ \bibnamefont {Oh}}, \bibinfo {author} {\bibfnamefont
  {D.}~\bibnamefont {Schneider}}, \bibinfo {author} {\bibfnamefont {R.~E.}\
  \bibnamefont {Lenski}}, \ and\ \bibinfo {author} {\bibfnamefont {J.~F.}\
  \bibnamefont {Kim}},\ }\href@noop {} {\bibfield  {journal} {\bibinfo
  {journal} {Nature}\ }\textbf {\bibinfo {volume} {461}},\ \bibinfo {pages}
  {1243} (\bibinfo {year} {2009})}\BibitemShut {NoStop}%
\bibitem [{\citenamefont {Bak}\ and\ \citenamefont {Sneppen}(1993)}]{bak1993}%
  \BibitemOpen
  \bibfield  {author} {\bibinfo {author} {\bibfnamefont {P.}~\bibnamefont
  {Bak}}\ and\ \bibinfo {author} {\bibfnamefont {K.}~\bibnamefont {Sneppen}},\
  }\href@noop {} {\bibfield  {journal} {\bibinfo  {journal} {Physical review
  letters}\ }\textbf {\bibinfo {volume} {71}},\ \bibinfo {pages} {4083}
  (\bibinfo {year} {1993})}\BibitemShut {NoStop}%
\bibitem [{\citenamefont {Harvey}\ \emph {et~al.}(1994)\citenamefont {Harvey},
  \citenamefont {May},\ and\ \citenamefont {Nee}}]{harvey1994}%
  \BibitemOpen
  \bibfield  {author} {\bibinfo {author} {\bibfnamefont {P.~H.}\ \bibnamefont
  {Harvey}}, \bibinfo {author} {\bibfnamefont {R.~M.}\ \bibnamefont {May}}, \
  and\ \bibinfo {author} {\bibfnamefont {S.}~\bibnamefont {Nee}},\ }\href@noop
  {} {\bibfield  {journal} {\bibinfo  {journal} {Evolution}\ }\textbf {\bibinfo
  {volume} {48}} (\bibinfo {year} {1994})}\BibitemShut {NoStop}%
\bibitem [{\citenamefont {Kubo}\ and\ \citenamefont {Y.}(1995)}]{phylo1}%
  \BibitemOpen
  \bibfield  {author} {\bibinfo {author} {\bibfnamefont {T.}~\bibnamefont
  {Kubo}}\ and\ \bibinfo {author} {\bibfnamefont {I.}~\bibnamefont {Y.}},\
  }\href@noop {} {\bibfield  {journal} {\bibinfo  {journal} {Evolution}\
  }\textbf {\bibinfo {volume} {49}} (\bibinfo {year} {1995})}\BibitemShut
  {NoStop}%
\bibitem [{\citenamefont {Avise}\ and\ \citenamefont
  {Wollenberg}(1997)}]{phylo2}%
  \BibitemOpen
  \bibfield  {author} {\bibinfo {author} {\bibfnamefont {J.}~\bibnamefont
  {Avise}}\ and\ \bibinfo {author} {\bibfnamefont {K.}~\bibnamefont
  {Wollenberg}},\ }\href@noop {} {\bibfield  {journal} {\bibinfo  {journal}
  {Proc. Natl Acad. Sci. USA}\ }\textbf {\bibinfo {volume} {94}} (\bibinfo
  {year} {1997})}\BibitemShut {NoStop}%
\bibitem [{\citenamefont {Stadler}(2008)}]{Stadler}%
  \BibitemOpen
  \bibfield  {author} {\bibinfo {author} {\bibfnamefont {T.}~\bibnamefont
  {Stadler}},\ }\href@noop {} {\bibfield  {journal} {\bibinfo  {journal}
  {Mathematical Biosciences}\ }\textbf {\bibinfo {volume} {216}} (\bibinfo
  {year} {2008})}\BibitemShut {NoStop}%
\bibitem [{\citenamefont {Houchmandzadeh}(2017)}]{houchmandzadeh2017}%
  \BibitemOpen
  \bibfield  {author} {\bibinfo {author} {\bibfnamefont {B.}~\bibnamefont
  {Houchmandzadeh}},\ }\href@noop {} {\bibfield  {journal} {\bibinfo  {journal}
  {Phys. Rev. E}\ }\textbf {\bibinfo {volume} {95}} (\bibinfo {year}
  {2017})}\BibitemShut {NoStop}%
\bibitem [{\citenamefont {Blythe}\ and\ \citenamefont {McKane}(2007)}]{Blythe}%
  \BibitemOpen
  \bibfield  {author} {\bibinfo {author} {\bibfnamefont {R.~A.}\ \bibnamefont
  {Blythe}}\ and\ \bibinfo {author} {\bibfnamefont {A.~J.}\ \bibnamefont
  {McKane}},\ }\href@noop {} {\bibfield  {journal} {\bibinfo  {journal} {J.
  Stat. Mech.: Theory Exp.}\ } (\bibinfo {year} {2007})}\BibitemShut {NoStop}%
\bibitem [{\citenamefont {Lambert}(2008)}]{lambert2008}%
  \BibitemOpen
  \bibfield  {author} {\bibinfo {author} {\bibfnamefont {A.}~\bibnamefont
  {Lambert}},\ }\href@noop {} {\bibfield  {journal} {\bibinfo  {journal}
  {Stochastic Models}\ }\textbf {\bibinfo {volume} {24}},\ \bibinfo {pages}
  {45} (\bibinfo {year} {2008})}\BibitemShut {NoStop}%
\bibitem [{\citenamefont {Lambert}(2010)}]{lambert2010}%
  \BibitemOpen
  \bibfield  {author} {\bibinfo {author} {\bibfnamefont {A.}~\bibnamefont
  {Lambert}},\ }\href@noop {} {\bibfield  {journal} {\bibinfo  {journal}
  {Journal of mathematical biology}\ }\textbf {\bibinfo {volume} {60}},\
  \bibinfo {pages} {469} (\bibinfo {year} {2010})}\BibitemShut {NoStop}%
\bibitem [{\citenamefont {Ewens}(2012)}]{ewens2012}%
  \BibitemOpen
  \bibfield  {author} {\bibinfo {author} {\bibfnamefont {W.~J.}\ \bibnamefont
  {Ewens}},\ }\href@noop {} {\emph {\bibinfo {title} {Mathematical population
  genetics 1: theoretical introduction}}},\ Vol.~\bibinfo {volume} {27}\
  (\bibinfo  {publisher} {Springer Science \& Business Media},\ \bibinfo {year}
  {2012})\BibitemShut {NoStop}%
\bibitem [{\citenamefont {Hubbell}(2001)}]{Hubbell}%
  \BibitemOpen
  \bibfield  {author} {\bibinfo {author} {\bibfnamefont {S.~P.}\ \bibnamefont
  {Hubbell}},\ }\href@noop {} {\emph {\bibinfo {title} {The unified neutral
  theory of biodiversity and biogeography (MPB-32)}}},\ Vol.~\bibinfo {volume}
  {32}\ (\bibinfo  {publisher} {Princeton University Press},\ \bibinfo {year}
  {2001})\BibitemShut {NoStop}%
\bibitem [{\citenamefont {Tilman}(2004)}]{Tilman2004}%
  \BibitemOpen
  \bibfield  {author} {\bibinfo {author} {\bibfnamefont {D.}~\bibnamefont
  {Tilman}},\ }\href@noop {} {\bibfield  {journal} {\bibinfo  {journal} {PNAS}\
  }\textbf {\bibinfo {volume} {101}} (\bibinfo {year} {2004})}\BibitemShut
  {NoStop}%
\bibitem [{\citenamefont {Azaele}\ \emph {et~al.}(2016)\citenamefont {Azaele},
  \citenamefont {Suweis}, \citenamefont {Grilli}, \citenamefont {Volkov},
  \citenamefont {Banavar},\ and\ \citenamefont {Maritan}}]{reviewNT}%
  \BibitemOpen
  \bibfield  {author} {\bibinfo {author} {\bibfnamefont {S.}~\bibnamefont
  {Azaele}}, \bibinfo {author} {\bibfnamefont {S.}~\bibnamefont {Suweis}},
  \bibinfo {author} {\bibfnamefont {J.}~\bibnamefont {Grilli}}, \bibinfo
  {author} {\bibfnamefont {I.}~\bibnamefont {Volkov}}, \bibinfo {author}
  {\bibfnamefont {J.}~\bibnamefont {Banavar}}, \ and\ \bibinfo {author}
  {\bibfnamefont {A.}~\bibnamefont {Maritan}},\ }\href@noop {} {\bibfield
  {journal} {\bibinfo  {journal} {Rev. Mod. Phys.}\ }\textbf {\bibinfo {volume}
  {88}} (\bibinfo {year} {2016})}\BibitemShut {NoStop}%
\bibitem [{\citenamefont {Houchmandzadeh}(2008)}]{houchmandzadeh2008}%
  \BibitemOpen
  \bibfield  {author} {\bibinfo {author} {\bibfnamefont {B.}~\bibnamefont
  {Houchmandzadeh}},\ }\href@noop {} {\bibfield  {journal} {\bibinfo  {journal}
  {Physical review letters}\ }\textbf {\bibinfo {volume} {101}},\ \bibinfo
  {pages} {078103} (\bibinfo {year} {2008})}\BibitemShut {NoStop}%
\bibitem [{\citenamefont {Houchmandzadeh}(2009)}]{houchmandzadeh2009}%
  \BibitemOpen
  \bibfield  {author} {\bibinfo {author} {\bibfnamefont {B.}~\bibnamefont
  {Houchmandzadeh}},\ }\href@noop {} {\bibfield  {journal} {\bibinfo  {journal}
  {Physical Review E}\ }\textbf {\bibinfo {volume} {80}},\ \bibinfo {pages}
  {051920} (\bibinfo {year} {2009})}\BibitemShut {NoStop}%
\bibitem [{\citenamefont {Volkov}\ \emph {et~al.}(2003)\citenamefont {Volkov},
  \citenamefont {Banavar}, \citenamefont {Hubbell},\ and\ \citenamefont
  {Maritan}}]{amos1}%
  \BibitemOpen
  \bibfield  {author} {\bibinfo {author} {\bibfnamefont {I.}~\bibnamefont
  {Volkov}}, \bibinfo {author} {\bibfnamefont {J.}~\bibnamefont {Banavar}},
  \bibinfo {author} {\bibfnamefont {S.~P.}\ \bibnamefont {Hubbell}}, \ and\
  \bibinfo {author} {\bibfnamefont {A.}~\bibnamefont {Maritan}},\ }\href@noop
  {} {\bibfield  {journal} {\bibinfo  {journal} {Nature}\ }\textbf {\bibinfo
  {volume} {424}} (\bibinfo {year} {2003})}\BibitemShut {NoStop}%
\bibitem [{\citenamefont {Volkov}\ \emph {et~al.}(2007)\citenamefont {Volkov},
  \citenamefont {Banavar}, \citenamefont {Hubbell},\ and\ \citenamefont
  {Maritan}}]{amos2}%
  \BibitemOpen
  \bibfield  {author} {\bibinfo {author} {\bibfnamefont {I.}~\bibnamefont
  {Volkov}}, \bibinfo {author} {\bibfnamefont {J.}~\bibnamefont {Banavar}},
  \bibinfo {author} {\bibfnamefont {S.~P.}\ \bibnamefont {Hubbell}}, \ and\
  \bibinfo {author} {\bibfnamefont {A.}~\bibnamefont {Maritan}},\ }\href@noop
  {} {\bibfield  {journal} {\bibinfo  {journal} {Nature}\ }\textbf {\bibinfo
  {volume} {450}} (\bibinfo {year} {2007})}\BibitemShut {NoStop}%
\bibitem [{\citenamefont {Kessler}\ \emph {et~al.}(2015)\citenamefont
  {Kessler}, \citenamefont {Suweis}, \citenamefont {Formentin},\ and\
  \citenamefont {Shnerb}}]{kessler2015}%
  \BibitemOpen
  \bibfield  {author} {\bibinfo {author} {\bibfnamefont {D.}~\bibnamefont
  {Kessler}}, \bibinfo {author} {\bibfnamefont {S.}~\bibnamefont {Suweis}},
  \bibinfo {author} {\bibfnamefont {M.}~\bibnamefont {Formentin}}, \ and\
  \bibinfo {author} {\bibfnamefont {N.~M.}\ \bibnamefont {Shnerb}},\
  }\href@noop {} {\bibfield  {journal} {\bibinfo  {journal} {Physical Review
  E}\ }\textbf {\bibinfo {volume} {92}},\ \bibinfo {pages} {022722} (\bibinfo
  {year} {2015})}\BibitemShut {NoStop}%
\bibitem [{\citenamefont {Hidalgo}\ \emph {et~al.}(2017)\citenamefont
  {Hidalgo}, \citenamefont {Suweis},\ and\ \citenamefont
  {Maritan}}]{hidalgo2017}%
  \BibitemOpen
  \bibfield  {author} {\bibinfo {author} {\bibfnamefont {J.}~\bibnamefont
  {Hidalgo}}, \bibinfo {author} {\bibfnamefont {S.}~\bibnamefont {Suweis}}, \
  and\ \bibinfo {author} {\bibfnamefont {A.}~\bibnamefont {Maritan}},\
  }\href@noop {} {\bibfield  {journal} {\bibinfo  {journal} {Journal of
  theoretical biology}\ }\textbf {\bibinfo {volume} {413}},\ \bibinfo {pages}
  {1} (\bibinfo {year} {2017})}\BibitemShut {NoStop}%
\bibitem [{\citenamefont {Mart{\'\i}n}\ \emph {et~al.}(2016)\citenamefont
  {Mart{\'\i}n}, \citenamefont {Hidalgo}, \citenamefont {de~Casas},\ and\
  \citenamefont {Mu{\~n}oz}}]{martin2016}%
  \BibitemOpen
  \bibfield  {author} {\bibinfo {author} {\bibfnamefont {P.~V.}\ \bibnamefont
  {Mart{\'\i}n}}, \bibinfo {author} {\bibfnamefont {J.}~\bibnamefont
  {Hidalgo}}, \bibinfo {author} {\bibfnamefont {R.~R.}\ \bibnamefont
  {de~Casas}}, \ and\ \bibinfo {author} {\bibfnamefont {M.~A.}\ \bibnamefont
  {Mu{\~n}oz}},\ }\href@noop {} {\bibfield  {journal} {\bibinfo  {journal}
  {PLoS computational biology}\ }\textbf {\bibinfo {volume} {12}},\ \bibinfo
  {pages} {e1005139} (\bibinfo {year} {2016})}\BibitemShut {NoStop}%
\bibitem [{\citenamefont {Chase}\ and\ \citenamefont
  {Leibold}(2003{\natexlab{a}})}]{Niche}%
  \BibitemOpen
  \bibfield  {author} {\bibinfo {author} {\bibfnamefont {J.~M.}\ \bibnamefont
  {Chase}}\ and\ \bibinfo {author} {\bibfnamefont {M.~A.}\ \bibnamefont
  {Leibold}},\ }\href@noop {} {\emph {\bibinfo {title} {Ecological niches:
  linking classical and contemporary approaches}}}\ (\bibinfo  {publisher}
  {University of Chicago Press},\ \bibinfo {year} {2003})\BibitemShut {NoStop}%
\bibitem [{\citenamefont {Chase}\ and\ \citenamefont
  {Leibold}(2003{\natexlab{b}})}]{chase2003ecological}%
  \BibitemOpen
  \bibfield  {author} {\bibinfo {author} {\bibfnamefont {J.~M.}\ \bibnamefont
  {Chase}}\ and\ \bibinfo {author} {\bibfnamefont {M.~A.}\ \bibnamefont
  {Leibold}},\ }\href@noop {} {\emph {\bibinfo {title} {Ecological niches:
  linking classical and contemporary approaches}}}\ (\bibinfo  {publisher}
  {University of Chicago Press},\ \bibinfo {year} {2003})\BibitemShut {NoStop}%
\bibitem [{\citenamefont {Kalyuzhny}\ \emph {et~al.}(2014)\citenamefont
  {Kalyuzhny}, \citenamefont {Seri}, \citenamefont {Chocron}, \citenamefont
  {Flather}, \citenamefont {Kadmon},\ and\ \citenamefont {Shnerb}}]{Nadav2014}%
  \BibitemOpen
  \bibfield  {author} {\bibinfo {author} {\bibfnamefont {M.}~\bibnamefont
  {Kalyuzhny}}, \bibinfo {author} {\bibfnamefont {E.}~\bibnamefont {Seri}},
  \bibinfo {author} {\bibfnamefont {R.}~\bibnamefont {Chocron}}, \bibinfo
  {author} {\bibfnamefont {C.~H.}\ \bibnamefont {Flather}}, \bibinfo {author}
  {\bibfnamefont {R.}~\bibnamefont {Kadmon}}, \ and\ \bibinfo {author}
  {\bibfnamefont {N.~M.}\ \bibnamefont {Shnerb}},\ }\href@noop {} {\bibfield
  {journal} {\bibinfo  {journal} {The American Naturalist}\ }\textbf {\bibinfo
  {volume} {184}} (\bibinfo {year} {2014})}\BibitemShut {NoStop}%
\bibitem [{\citenamefont {Pigolotti}\ \emph {et~al.}(2010)\citenamefont
  {Pigolotti}, \citenamefont {Lopez}, \citenamefont {Hernandez-Garcia},\ and\
  \citenamefont {Adersen}}]{pigolotti2010}%
  \BibitemOpen
  \bibfield  {author} {\bibinfo {author} {\bibfnamefont {S.}~\bibnamefont
  {Pigolotti}}, \bibinfo {author} {\bibfnamefont {C.}~\bibnamefont {Lopez}},
  \bibinfo {author} {\bibfnamefont {E.}~\bibnamefont {Hernandez-Garcia}}, \
  and\ \bibinfo {author} {\bibfnamefont {K.~H.}\ \bibnamefont {Adersen}},\
  }\href@noop {} {\bibfield  {journal} {\bibinfo  {journal} {Theor Ecol}\
  }\textbf {\bibinfo {volume} {89}} (\bibinfo {year} {2010})}\BibitemShut
  {NoStop}%
\bibitem [{\citenamefont {Hardin}(1960)}]{hardin1960}%
  \BibitemOpen
  \bibfield  {author} {\bibinfo {author} {\bibfnamefont {G.}~\bibnamefont
  {Hardin}},\ }\href@noop {} {\bibfield  {journal} {\bibinfo  {journal}
  {science}\ }\textbf {\bibinfo {volume} {131}},\ \bibinfo {pages} {1292}
  (\bibinfo {year} {1960})}\BibitemShut {NoStop}%
\bibitem [{\citenamefont {Harpole}\ and\ \citenamefont
  {Tilman}(2007)}]{harpole2007}%
  \BibitemOpen
  \bibfield  {author} {\bibinfo {author} {\bibfnamefont {W.~S.}\ \bibnamefont
  {Harpole}}\ and\ \bibinfo {author} {\bibfnamefont {D.}~\bibnamefont
  {Tilman}},\ }\href@noop {} {\bibfield  {journal} {\bibinfo  {journal}
  {Nature}\ }\textbf {\bibinfo {volume} {446}},\ \bibinfo {pages} {791}
  (\bibinfo {year} {2007})}\BibitemShut {NoStop}%
\bibitem [{\citenamefont {Leibold}\ and\ \citenamefont
  {McPeek}(2006)}]{leibold2006}%
  \BibitemOpen
  \bibfield  {author} {\bibinfo {author} {\bibfnamefont {M.~A.}\ \bibnamefont
  {Leibold}}\ and\ \bibinfo {author} {\bibfnamefont {M.~A.}\ \bibnamefont
  {McPeek}},\ }\href@noop {} {\bibfield  {journal} {\bibinfo  {journal}
  {Ecology}\ }\textbf {\bibinfo {volume} {87}},\ \bibinfo {pages} {1399}
  (\bibinfo {year} {2006})}\BibitemShut {NoStop}%
\bibitem [{\citenamefont {Chisholm}\ and\ \citenamefont
  {Pacala}(2010)}]{chisholm2010niche}%
  \BibitemOpen
  \bibfield  {author} {\bibinfo {author} {\bibfnamefont {R.~A.}\ \bibnamefont
  {Chisholm}}\ and\ \bibinfo {author} {\bibfnamefont {S.~W.}\ \bibnamefont
  {Pacala}},\ }\href@noop {} {\bibfield  {journal} {\bibinfo  {journal}
  {Proceedings of the National Academy of Sciences}\ ,\ \bibinfo {pages}
  {201009387}} (\bibinfo {year} {2010})}\BibitemShut {NoStop}%
\bibitem [{\citenamefont {Haegeman}\ and\ \citenamefont
  {Loreau}(2011)}]{haegeman2011}%
  \BibitemOpen
  \bibfield  {author} {\bibinfo {author} {\bibfnamefont {B.}~\bibnamefont
  {Haegeman}}\ and\ \bibinfo {author} {\bibfnamefont {M.}~\bibnamefont
  {Loreau}},\ }\href@noop {} {\bibfield  {journal} {\bibinfo  {journal}
  {Journal of Theoretical Biology}\ }\textbf {\bibinfo {volume} {269}},\
  \bibinfo {pages} {150} (\bibinfo {year} {2011})}\BibitemShut {NoStop}%
\bibitem [{\citenamefont {Fisher}\ and\ \citenamefont
  {Mehta}(2014)}]{fisher2014}%
  \BibitemOpen
  \bibfield  {author} {\bibinfo {author} {\bibfnamefont {C.~K.}\ \bibnamefont
  {Fisher}}\ and\ \bibinfo {author} {\bibfnamefont {P.}~\bibnamefont {Mehta}},\
  }\href@noop {} {\bibfield  {journal} {\bibinfo  {journal} {Proceedings of the
  National Academy of Sciences}\ }\textbf {\bibinfo {volume} {111}},\ \bibinfo
  {pages} {13111} (\bibinfo {year} {2014})}\BibitemShut {NoStop}%
\bibitem [{\citenamefont {Lafuerza}\ and\ \citenamefont
  {McKane}(2016)}]{lafuerza}%
  \BibitemOpen
  \bibfield  {author} {\bibinfo {author} {\bibfnamefont {L.~F.}\ \bibnamefont
  {Lafuerza}}\ and\ \bibinfo {author} {\bibfnamefont {A.~J.}\ \bibnamefont
  {McKane}},\ }\href@noop {} {\bibfield  {journal} {\bibinfo  {journal}
  {Physical Review E}\ }\textbf {\bibinfo {volume} {93}},\ \bibinfo {pages}
  {032121} (\bibinfo {year} {2016})}\BibitemShut {NoStop}%
\bibitem [{\citenamefont {Gillespie}(2007)}]{Gillespie}%
  \BibitemOpen
  \bibfield  {author} {\bibinfo {author} {\bibfnamefont {D.~T.}\ \bibnamefont
  {Gillespie}},\ }\href@noop {} {\bibfield  {journal} {\bibinfo  {journal} {Ann
  Rev Phys Chem}\ }\textbf {\bibinfo {volume} {58}} (\bibinfo {year}
  {2007})}\BibitemShut {NoStop}%
\bibitem [{\citenamefont {Van~Kampen}(2007)}]{VanKampen}%
  \BibitemOpen
  \bibfield  {author} {\bibinfo {author} {\bibfnamefont {N.~G.}\ \bibnamefont
  {Van~Kampen}},\ }\href@noop {} {\emph {\bibinfo {title} {Stochastic processes
  in physics and chemistry}}}\ (\bibinfo  {publisher} {Elsevier},\ \bibinfo
  {year} {2007})\BibitemShut {NoStop}%
\bibitem [{\citenamefont {Al~Hammal}\ \emph {et~al.}(2005)\citenamefont
  {Al~Hammal}, \citenamefont {Chat{\'e}}, \citenamefont {Dornic},\ and\
  \citenamefont {Munoz}}]{alhammal2005}%
  \BibitemOpen
  \bibfield  {author} {\bibinfo {author} {\bibfnamefont {O.}~\bibnamefont
  {Al~Hammal}}, \bibinfo {author} {\bibfnamefont {H.}~\bibnamefont
  {Chat{\'e}}}, \bibinfo {author} {\bibfnamefont {I.}~\bibnamefont {Dornic}}, \
  and\ \bibinfo {author} {\bibfnamefont {M.~A.}\ \bibnamefont {Munoz}},\
  }\href@noop {} {\bibfield  {journal} {\bibinfo  {journal} {Physical review
  letters}\ }\textbf {\bibinfo {volume} {94}},\ \bibinfo {pages} {230601}
  (\bibinfo {year} {2005})}\BibitemShut {NoStop}%
\bibitem [{\citenamefont {Gardiner}(2009)}]{Gardiner2009}%
  \BibitemOpen
  \bibfield  {author} {\bibinfo {author} {\bibfnamefont {C.}~\bibnamefont
  {Gardiner}},\ }\href@noop {} {\emph {\bibinfo {title} {Stochastic
  methods}}},\ Vol.~\bibinfo {volume} {4}\ (\bibinfo  {publisher} {springer
  Berlin},\ \bibinfo {year} {2009})\BibitemShut {NoStop}%
\bibitem [{\citenamefont {Di~Patti}\ \emph {et~al.}(2011)\citenamefont
  {Di~Patti}, \citenamefont {Azaele}, \citenamefont {Banavar},\ and\
  \citenamefont {Maritan}}]{diPatti2011}%
  \BibitemOpen
  \bibfield  {author} {\bibinfo {author} {\bibfnamefont {F.}~\bibnamefont
  {Di~Patti}}, \bibinfo {author} {\bibfnamefont {S.}~\bibnamefont {Azaele}},
  \bibinfo {author} {\bibfnamefont {J.~R.}\ \bibnamefont {Banavar}}, \ and\
  \bibinfo {author} {\bibfnamefont {A.}~\bibnamefont {Maritan}},\ }\href@noop
  {} {\bibfield  {journal} {\bibinfo  {journal} {Physical Review E}\ }\textbf
  {\bibinfo {volume} {83}},\ \bibinfo {pages} {010102} (\bibinfo {year}
  {2011})}\BibitemShut {NoStop}%
\bibitem [{\citenamefont {Rogers}\ and\ \citenamefont
  {Rossberg}(2012{\natexlab{b}})}]{McKane0}%
  \BibitemOpen
  \bibfield  {author} {\bibinfo {author} {\bibfnamefont {A.~J.}\ \bibnamefont
  {Rogers}, \bibfnamefont {T.~McKane}}\ and\ \bibinfo {author} {\bibfnamefont
  {A.~G.}\ \bibnamefont {Rossberg}},\ }\href@noop {} {\bibfield  {journal}
  {\bibinfo  {journal} {Phys. Bio.}\ }\textbf {\bibinfo {volume} {9}} (\bibinfo
  {year} {2012}{\natexlab{b}})}\BibitemShut {NoStop}%
\bibitem [{\citenamefont {Rossberg}\ and\ \citenamefont
  {Rogers}(2013)}]{McKaneLTT}%
  \BibitemOpen
  \bibfield  {author} {\bibinfo {author} {\bibfnamefont {A.~G.}\ \bibnamefont
  {Rossberg}}\ and\ \bibinfo {author} {\bibfnamefont {A.~J.}\ \bibnamefont
  {Rogers}, \bibfnamefont {T.~McKane}},\ }\href@noop {} {\bibfield  {journal}
  {\bibinfo  {journal} {Proc Biol Sci.}\ }\textbf {\bibinfo {volume} {280}}
  (\bibinfo {year} {2013})}\BibitemShut {NoStop}%
\end{thebibliography}%

\newpage
\end{document}